\documentclass{aa}
\usepackage{graphicx}
\usepackage{txfonts}
\begin{document}
  \title{
    Extrasolar planets and brown dwarfs around A-F type stars
    \thanks{Based on observations made with the {\small ELODIE} spectrograph at
      the Observatoire de Haute-Provence (CNRS, France).}
  }
  \subtitle{II. A planet found with ELODIE around the F6V star HD\,33564.}
  
  \author{
    F. Galland \inst{1,2}
    \and
    A.-M. Lagrange \inst{1}
    \and
    S. Udry \inst{2}
    \and
    A. Chelli \inst{1}
    \and
    F. Pepe \inst{2}
    \and
    J.-L. Beuzit \inst{1}
    \and
    M. Mayor \inst{2}
  }
  
  \offprints{
    F. Galland,\\
    \email{Franck.Galland@obs.ujf-grenoble.fr}
  }

  \institute{
    Laboratoire d'Astrophysique de l'Observatoire de Grenoble,
    Universit\'e Joseph Fourier, BP 53, 38041 Grenoble, France
    \and
    Observatoire de Gen\`eve, 51 Ch. des Maillettes, 1290 Sauverny, Switzerland
  }
  
  \date{Received 3 June 2005 / Accepted 3 September 2005}

  \abstract{

    We present here the detection of a planet orbiting around the F6V
    star HD\,33564. The radial velocity measurements, obtained with the
    {\small ELODIE} echelle spectrograph at the Haute-Provence Observatory,
    show a variation with a period of 388 days. Assuming a primary
    mass of 1.25~M$_{\odot}$, the best Keplerian fit to the data leads
    to a minimum mass of 9.1~M$_{\rm Jup}$ for the companion.

  }
  
  \maketitle
  
  \section{Introduction}

  Radial velocity surveys have lead to the detection of more than 150
  planets during the past decade \footnote{A comprehensive list of
  known exoplanets is available at
  http://www.obspm.fr/encycl/cat1.html}.  So far, they have been
  limited to solar and later-type stars ($\ga$ F7), as it was thought
  that planets around more massive stars were not accessible to radial
  velocity techniques. They present a small number of stellar lines,
  usually broadened and blended by stellar rotation. However, we
  recently showed (\cite{Galland05a}, Paper\,I) that with a new radial
  velocity measurement method, it was possible to detect planets even
  around early A-type stars. Finding planets around massive stars is
  of importance, as this will allow us to test planetary formation and
  evolution processes around a larger variety of objects.
  
  We have started a radial velocity survey dedicated to the search for
  extrasolar planets and brown dwarfs around a volume-limited sample
  of A-F Main-Sequence stars i) with the {\small ELODIE} fiber-fed
  echelle spectrograph (\cite{Baranne96}) mounted on the 1.93-m
  telescope at the Observatoire de Haute-Provence (CNRS, France) in
  the northern hemisphere, and ii) with the {\small HARPS}
  spectrograph (\cite{Pepe02}) installed on the 3.6-m ESO telescope at
  La Silla Observatory (ESO, Chile) in the southern hemisphere. The
  method, achieved detection limits as well as estimates of the
  minimum detectable masses are described in Paper\,I. We present here
  the detection of a planet around one of the objects surveyed
  with {\small ELODIE}, HD\,33564. Section~2 provides the stellar
  properties, the results of the radial velocity fit and related
  comments. In Section~3, we rule out other possible origins of the
  observed radial velocity variations. Finally, we discuss the status
  of this system in the last section.

  \section{A planet around HD\,33564}

  \subsection{Stellar properties}

  HD\,33564 (HIP 25110, HR 1686) is located at 21.0\,pc from the Sun
  (ESA 1997).  Stellar parameters such as mass
  $M_1$\,=\,$1.25_{-0.04}^{+0.03}$\,M$_{\odot}$, age
  $3.0_{-0.3}^{+0.6}$\,Gyr, metallicity [Fe/H]\,=\,-0.12 are taken
  from \cite{Gen-Cop-Surv-Sol-Neigh}.  Rotational velocity
  $v\sin{i}$\,=\,12\,km\,s$^{\rm -1}$, effective temperature
  $T_{\rm eff}$\,=\,6250\,K, and surface gravity $\log{g}$\,=\,4.0 are
  taken from \cite{Acke04}, see
  Table\,\ref{Table_hd33564_stelpar}. These values are consistent with
  an F6V spectral type, commonly attributed to this star as e.g. in
  the HIPPARCOS catalogue (ESA 1997) or in the Bright Star Catalogue
  (\cite{Hoff91}).

  An infrared excess has been detected at 60\,$\mu m$ toward HD\,33564
  (DM\,+79\,169) with IRAS (\cite{Aumann85}, \cite{Patten91}), which
  suggested that this star could be a Vega-like star, surrounded by a
  cold (65\,K) dusty circumstellar disk, remnant of the formation of
  the system. Indeed, some Spitzer results show that this infrared
  excess is due to a background galaxy and not to a circumstellar disk
  around HD\,33564 (Bryden et al. 2006).

  \begin{table}[t!]
    \caption{HD\,33564 stellar properties. Photometric and
      astrometric data are extracted from the HIPPARCOS catalogue
    (ESA 1997); spectroscopic data are from
      \cite{Gen-Cop-Surv-Sol-Neigh} and \cite{Acke04}. 
    }
    \label{Table_hd33564_stelpar}
    \begin{center}
      \begin{tabular}{l l c}
        \hline
	\hline
        Parameter       &        & HD\,33564  \\
	\hline
        Spectral Type   &        & F6V \\
        $v\sin{i}$      & [km\,s$^{\rm -1}$] & 12  \\
        V               &        & 5.08 \\
        B-V             &        & 0.506 \\
        $\pi$           & [mas]  & $47.66 \pm 0.52$  \\
        Distance        & [pc]   & 21.0 \\
        $M_V$           &        & 3.47 \\
	\,[Fe/H]         &        & $-$ 0.12 \\
        $T_{\rm eff}$       & [K]    & 6250  \\
        log $g$         &        & 4.0  \\
        $M_1$         & [$M_\odot$] &  $1.25_{-0.04}^{+0.03}$ \\
	Age             & [Gyr]   & $3.0_{-0.3}^{+0.6}$  \\
        \hline
      \end{tabular}
    \end{center}
  \end{table}

  \subsection{Radial velocity measurements}

  By April 2005, 15 spectra have been acquired with {\small ELODIE}
  over a time span of 417 days. The wavelength range of the spectra is
  3850-6800\,$\AA$. The typical exposure time was 15 min, leading to
  $\mathrm{S/N}$ equal to $\sim$\,150. The exposures were performed
  with the simultaneous-thorium mode to follow and correct for the
  local astroclimatic drift of the instrument. The radial velocities
  have been measured using the method described in \cite{Chelli00} and
  Paper\,I. They are displayed in Fig~\ref{hd33564_dvft}.  The
  uncertainty is 11\,m\,s$^{\rm -1}$ on average. It is consistent with
  the value of 10\,m\,s$^{\rm -1}$ obtained from simulations in
  Paper\,I by applying the relation between the radial velocity
  uncertainties and $v\sin{i}$ to HD\,33564, with $\mathrm{S/N}$
  values equal to 150.

  \subsection{Orbital parameters}

\begin{figure}[t!]
    \centering
    \includegraphics[width=1.\hsize]{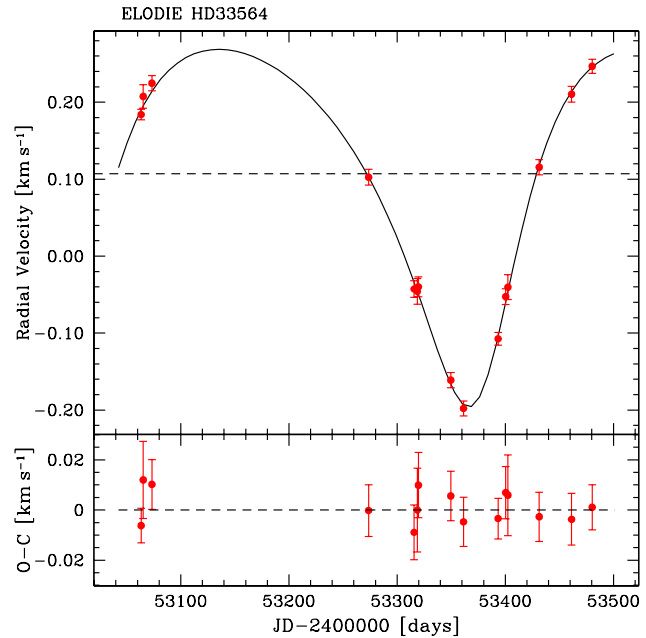}
    \caption{ Top: {\small ELODIE} radial velocities and orbital solution for
    HD\,33564. Bottom: Residuals to the fitted orbital solution.}
    \label{hd33564_dvft}
  \end{figure}

  The amplitude of the radial velocity variations is much larger than
  the uncertainties. The orbital parameters derived from the best
  Keplerian solution (Fig.\,\ref{hd33564_dvft}) are given in
  Table\,\ref{Table_hd33564_rvpar}. The residuals dispersion is
  7\,m\,s$^{\rm -1}$. The eccentricity is 0.34. Assuming a primary
  mass of 1.25\,M$_\odot$, the companion mass falls in the planetary
  domain, with a minimum mass of 9.1\,M$_{\rm Jup}$. The orbital period is
  388\,days (separation of 1.1\,AU).

  Note that the maximum of the radial velocity variations could not be
  observed. The orbital period is close to one year and a long time
  will still be necessary to be able to cover the whole phase.  To
  quantify the resulting uncertainty on $m_2\sin{i}$, we tried several
  fits fixing different values for the eccentricity. We find that
  residuals are acceptable with values of this eccentricity between
  0.28 and 0.40. The resulting minimum mass ranges then from 8.7 to
  9.6\,M$_{\rm Jup}$, hence an uncertainty of 6\,\% on the minimum mass
  corresponding to the best solution (given the primary mass).

  Besides, this star was also observed with the {\small CORAVEL}
  spectrograph. The obtained radial velocities are constant over
  4~years given uncertainties of 500\,m\,s$^{\rm -1}$ on average. The
  radial velocity variations obtained with {\small ELODIE} have a
  semi-amplitude lower than 250\,m\,s$^{\rm -1}$, not detectable with
  {\small CORAVEL}.

 \begin{table}[t!]
    \caption{{\small ELODIE} best orbital solution for HD\,33564.}
    \label{Table_hd33564_rvpar}
    \begin{center}
      \begin{tabular}{l l c}
	\hline
        \hline
        Parameter     &                          & HD\,33564 \\
        \hline
        P             & [days]                   & $388   \pm 3$    \\
        T             & [JD-2400000]             & $52603 \pm 8$    \\
        e             &                          & $0.34  \pm 0.02$  \\
        $\gamma$      & [km\,s$^{\rm -1}$]       & $0.107 \pm 0.006$ \\
        $\omega$      & [deg]                    & $205   \pm 4$     \\
        K             & [km\,s$^{\rm -1}$]       & $0.232 \pm 0.005$ \\
        N$_{\rm meas}$    &                          & 15    \\
        $\sigma(O-C)$ & [m\,s$^{\rm -1}$]        & 6.7   \\
        \hline
        $a_1\sin{i}$  & [$10^{-3}$~AU]           & 7.8   \\
        f(m)          & [$10^{-7} M_{\odot}$]    & 4.21   \\
        $M_1$         & [$M_{\odot}$]            & 1.25   \\
        $m_2\sin{i}$  & [M$_{\rm Jup}$]              & 9.1    \\
        $a$           & [AU]                     & 1.1    \\
        \hline
      \end{tabular}
    \end{center}
  \end{table}

  \begin{figure}[t!]
    \centering
     \includegraphics[bb= 70 241 682 709,width=1.\hsize]{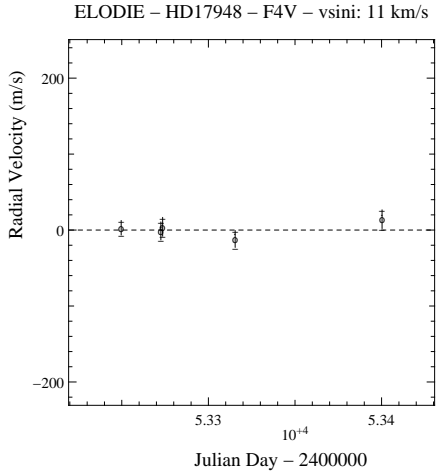}
     \caption{{\small ELODIE} radial velocity data for HD\,17948, a
     star constant in radial velocity (dispersion of 8.4\,m\,s$^{\rm
     -1}$), similar and close to HD\,33564. The vertical scale is
     identical to the one in Fig.\,\ref{hd33564_dvft}.}
    \label{hd17948_vr}
  \end{figure}

 \begin{figure*}[t!]
   \centering
    \includegraphics[width=0.32\hsize]{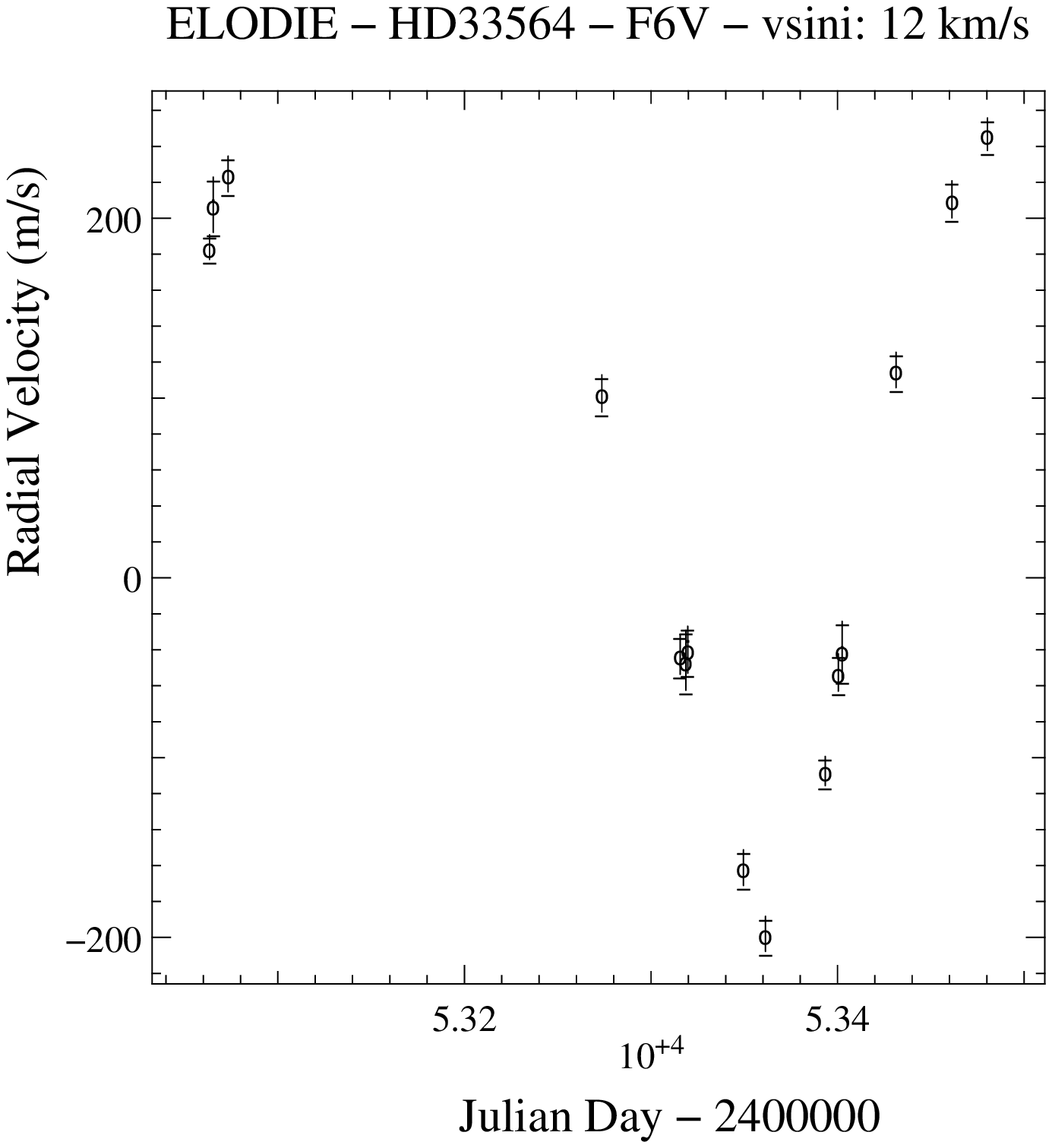}
    \includegraphics[width=0.32\hsize]{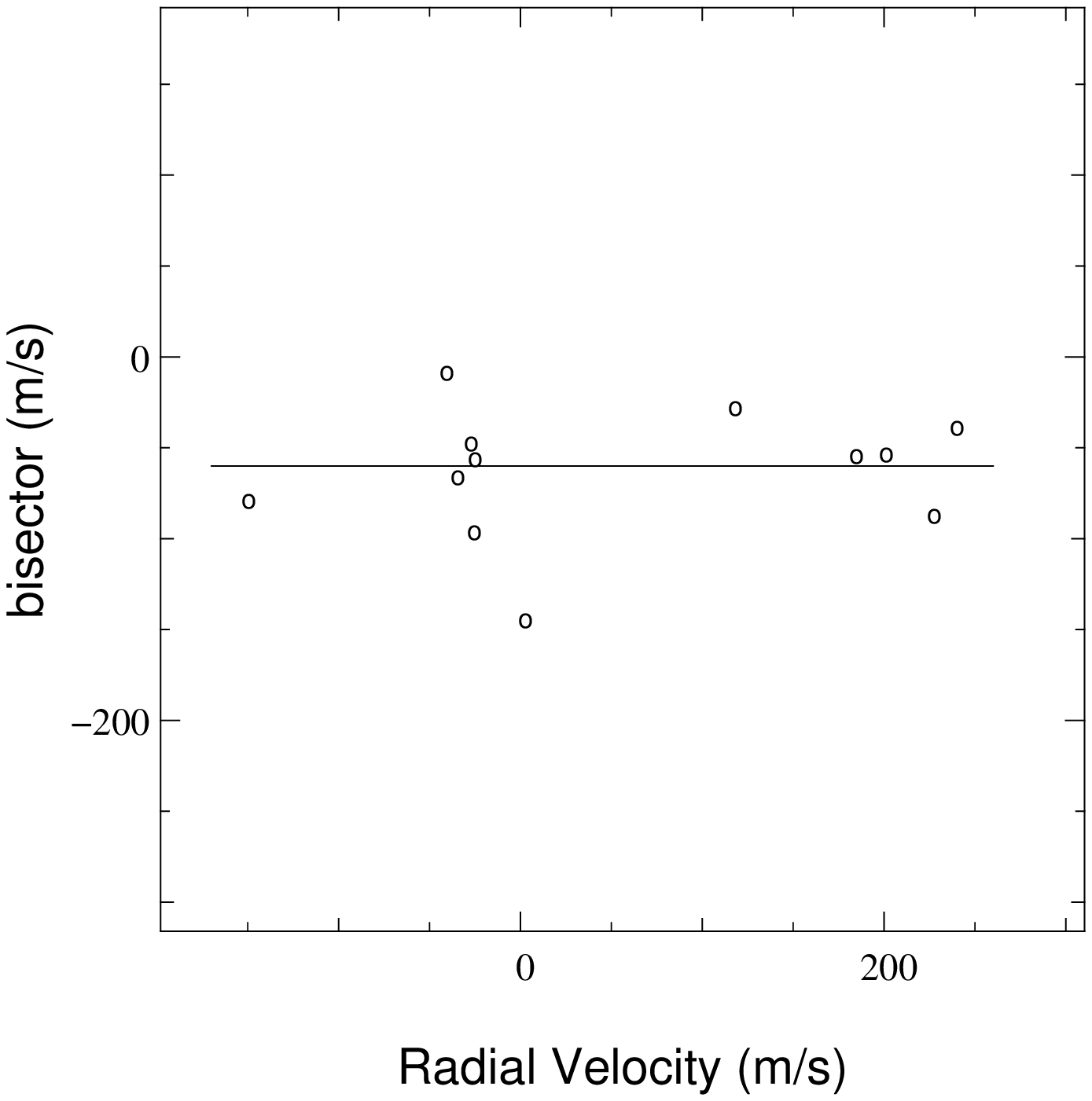}
    \includegraphics[width=0.32\hsize]{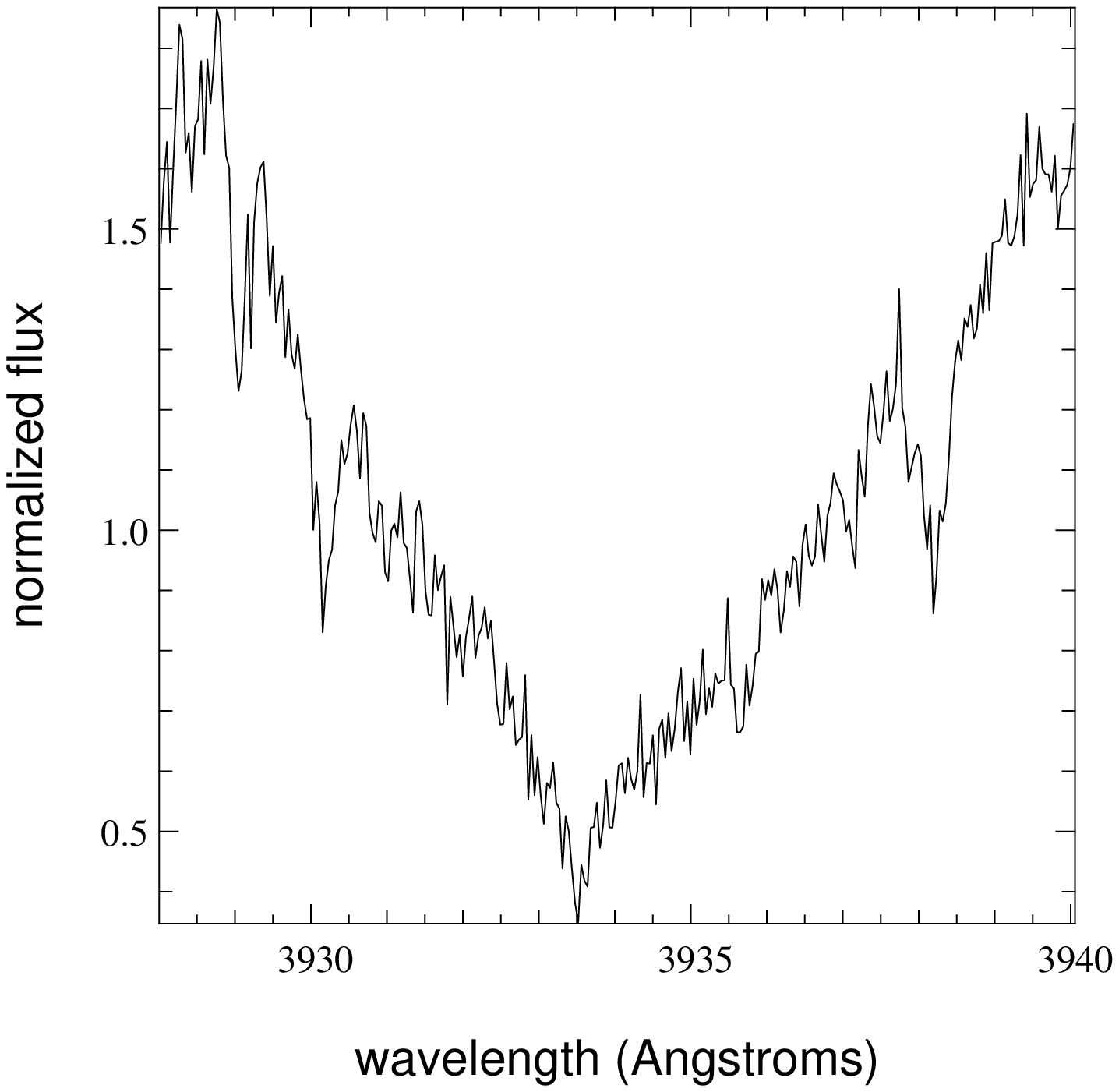}
    \includegraphics[width=0.32\hsize]{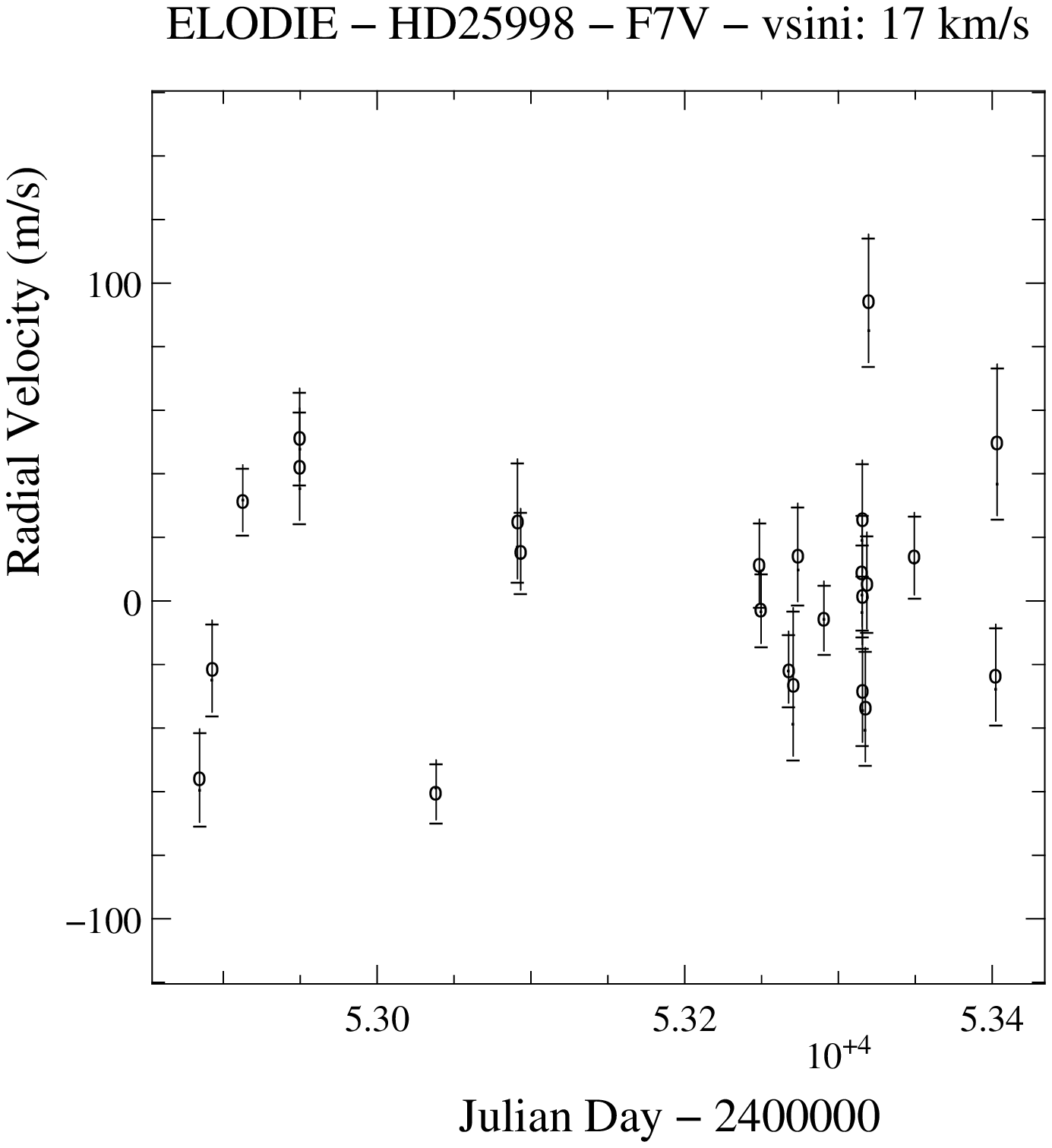}
    \includegraphics[width=0.32\hsize]{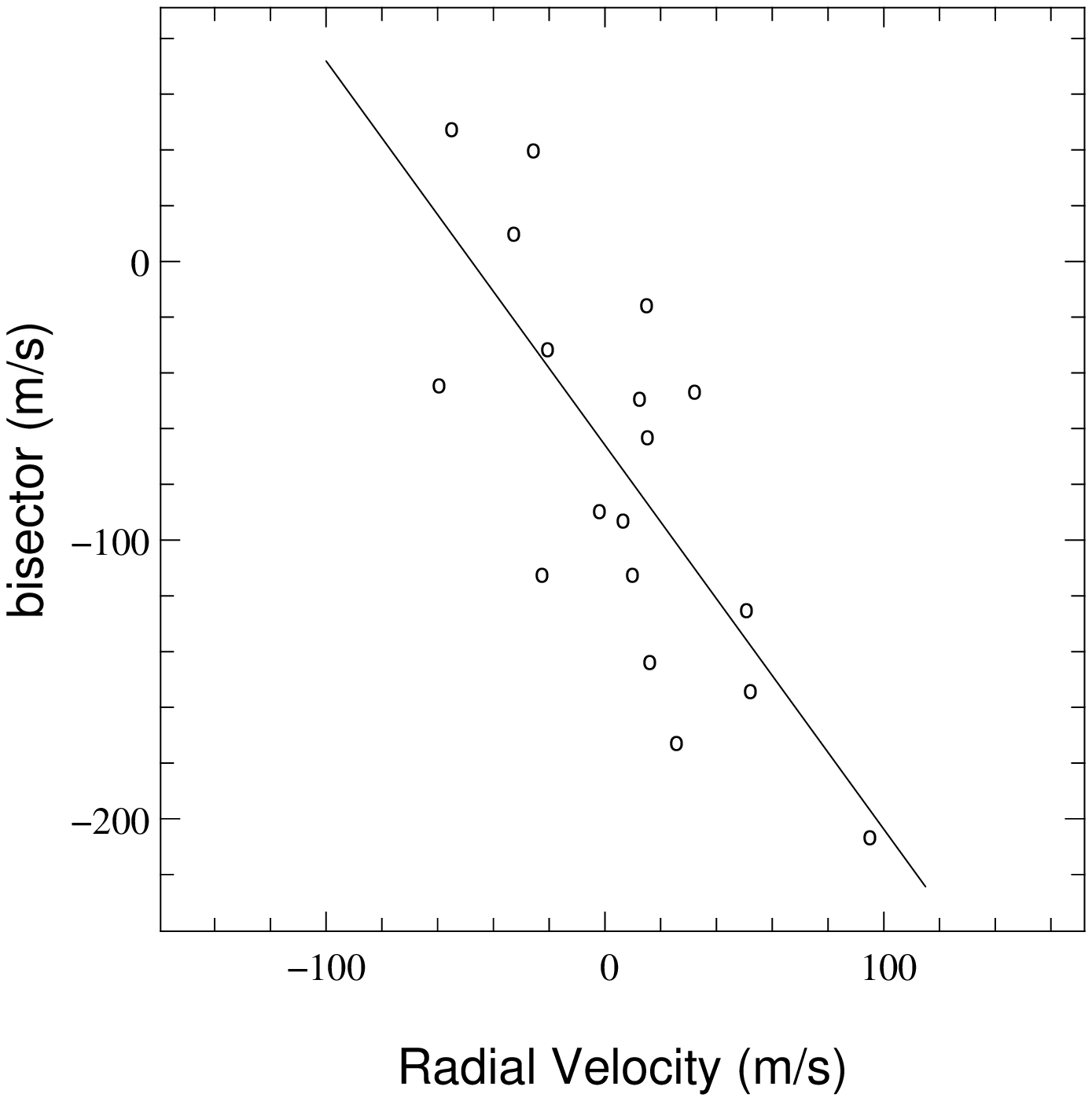}
    \includegraphics[width=0.32\hsize]{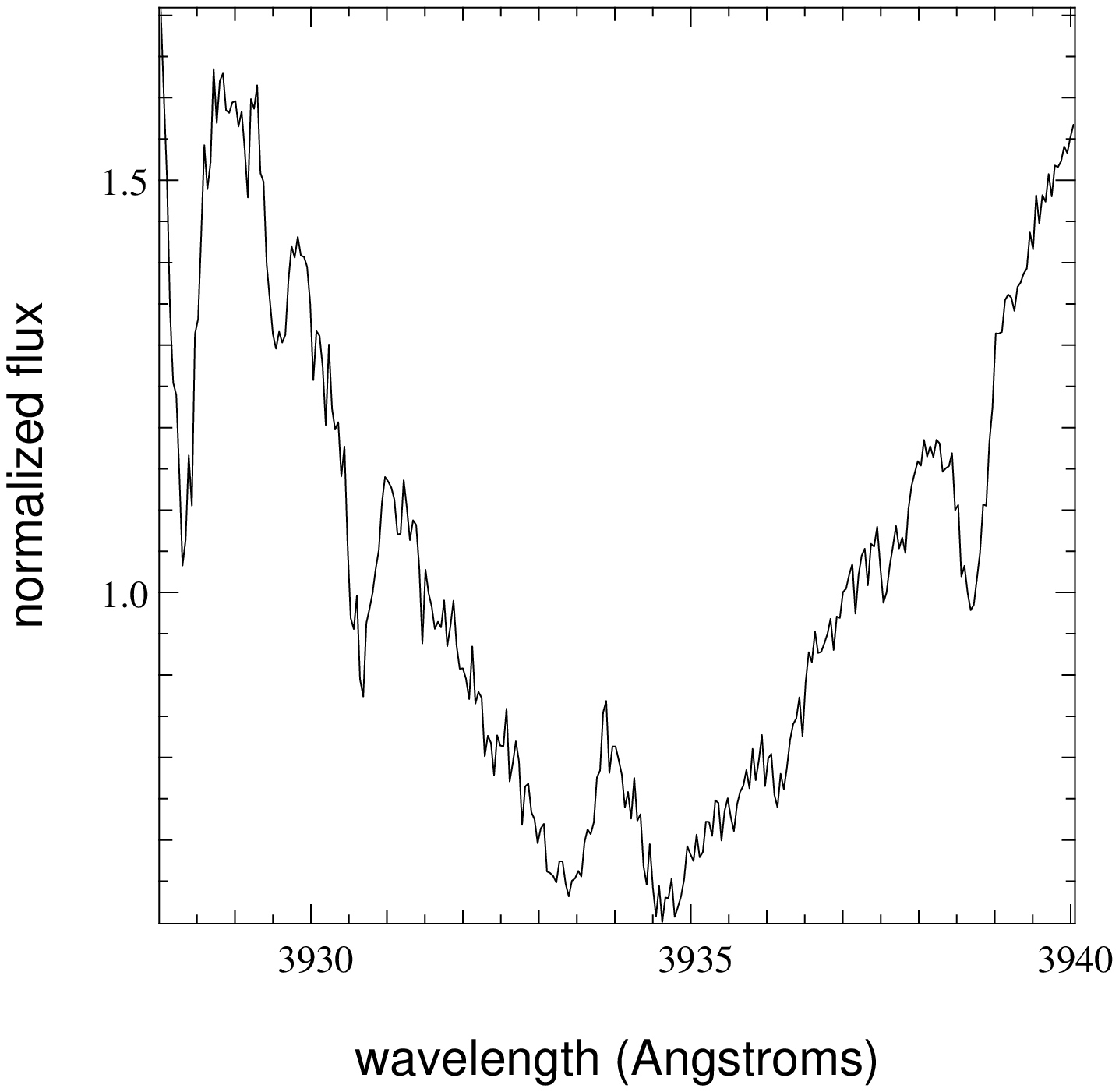}
    \caption{ Left: radial velocity measurements of HD\,33564 (top)
      and HD\,25998 (bottom): they are significantly variable.  Center:
      the line bisector (see text) is not variable for HD\,33564,
      while it is variable for HD\,25998, in a similar range, and
      linearly related to the radial velocities. This is typical of
      the presence of spots, induced by stellar activity
      (\cite{Queloz2001a} 2001a).  Right: no emission in the Ca K line in
      the case of HD\,33564, emission in the case of HD\,25998.
    }
    \label{bisect_Ca}
  \end{figure*}

 \section{Ruling out other origins of the variations}

  \subsection{A single star}

  HD\,33564 is referenced as a double star, but the high difference
  between the proper motions of the two components indicates that they
  are probably not bound.  Besides, the separation between the two
  components is larger than 6~arcsec, so that the flux from the
  secondary does not enter into the fiber of the spectrograph, even in
  bad seeing conditions. Hence, the spectra are not polluted by the
  secondary spectra.  In both cases, the presence of the secondary is
  not responsible for the observed radial velocity variations of the considered
  component, HD\,33564\,A.

  \subsection{A similar star constant in radial velocity}

  To rule out any possibility of artifact linked with the Earth
  orbital motion, we show in Fig.\,\ref{hd17948_vr} the radial
  velocities of a similar star, close to HD\,33564, but constant 
  within the present level of uncertainties: HD\,17948 is an F4V star with
  $v\sin{i}$\,=\,11\,km\,s$^{\rm -1}$. It also belongs to our
  {\small ELODIE} survey.
  By April~2005, 5 spectra have been gathered for this star, with
  $\mathrm{S/N}$ equal to 136 on average.  The typical uncertainty is
  10\,m\,s$^{\rm -1}$ comparable to the one obtained for
  HD\,33564. The observed radial velocity dispersion of 8.4 m\,s$^{\rm
  -1}$ shows that star is constant over the 150~days of the
  measurement span.

  \subsection{An activity-quiet star}

  We show here that activity is not responsible for the radial
  velocity variations of HD\,33564.  First, the bisector shape of the
  lines is estimated on the cross-correlation function (inverse
  bisector slope) in the same way as in \cite{Queloz2001a} (2001a). It is
  displayed in Fig\,\ref{bisect_Ca}, together with the one of a
  known active star (\cite{Gray03} 2003) included in our {\small ELODIE}
  survey, HD\,25998 (F7V, $v\sin{i}$\,=\,17\,km\,s$^{\rm -1}$), for
  comparison. It is flat i.e. there is no correlation between the
  radial velocities and the shape of the lines for HD\,33564. On the
  other hand, it varies linearly with the radial velocity in the case
  of HD\,25998 (the slope value is -1.4), which is typical of spots
  induced by stellar activity (see \cite{Queloz2001a} 2001a for further
  details).

  Moreover, no emission in the core of the Ca\,II lines is observed
  (see Fig\,\ref{bisect_Ca} for the case of the Ca\,II\,K~line). The
  chromospheric flux in the Ca\,II K and H~lines of HD\,33564 has been
  measured by e.g. \cite{Gray03} (2003), who gives an activity index
  $\log(R_{HK}')$\,=\,$-4.95$, leading to the classification of
  HD\,33564 as an inactive star. For comparison, the value of
  $\log(R_{HK}')$ is $-4.47$ in the case of the active star HD\,25998.
  Hence, contrary to HD\,25998, activity is not responsible for the
  radial velocity variations observed for HD\,33564. This is also
  confirmed by the fact that the period of the radial velocity
  variations is much larger than the rotational period of HD\,33564
  (less than 7~days given its value of $v\sin{i}$).

  \section{Status of the HD\,33564 system}

  So far 16 stars with spectral type between F7 and F9 have been
  reported as hosting possible planets detected through radial
  velocities.  HD\,33564 is, to our knowledge, the earliest spectral
  type star around which a companion with $m_2\sin{i}$ in the
  planetary domain has been announced.  The reported $m_2\sin{i}$\,'s
  range between 0.4 and 11\,M$_{\rm Jup}$.  Four objects, namely
  HD\,23596, HD\,89744, HD\,114762 and HD\,136118, are found with
  $m_2\sin{i}$\,$\geq$\,5\,M$_{\rm Jup}$ i.e. 8.0, 7.2, 11.0 and
  11.9\,M$_{\rm Jup}$, respectively. The corresponding estimated
  planet-star separations are 2.7, 0.88, 0.37 and 2.3\,AU,
  respectively.  HD\,33564, with $m_2\sin{i}$\,=\,9.1\,M$_{\rm Jup}$ and
  a\,=\,1.1\,AU, falls in a similar (mass, separation) domain. Knowing
  the inclination of these systems would be very important for
  constraining planet formation models. If their true masses are found
  to be planetary, one may wonder how such massive planets could form
  and stand so close to their parent stars. Whether migration has
  occurred or whether the planets have been formed by gravitational
  collapse rather than by a scenario including accretion onto a solid
  core are
  possibilities that need to be investigated.  The proposed
  idea that ``the more massive the star, the more massive the
  planets hosted'' is also very interesting. Studies on lower mass
  stars (\cite{Lin05}) show such a trend, does it exist for A-F type
  stars?
 
  Besides, we may emphasize the low value of the metallicity for
  HD\,33564, which is not common among stars hosting planets. There is
  however no metallicity trend for the early-type stars with massive
  planets mentioned above.

  \section{Conclusions}

  We have presented here the first detection of a planet around one of
  the objects surveyed in our {\small ELODIE} program, HD\,33564, an
  F6V star with $v\sin{i}$\,=\,12\,km\,$s ^{\rm -1}$. The best
  Keplerian solution derived from the radial velocity measurements
  leads to a minimum mass of 9.1\,M$_{\rm Jup}$ and a period of 388~days,
  hence a separation of 1.1\,AU.
  
  This result is a first step toward the extension of the study of
  planet and brown-dwarf formation processes to stars earlier than
  F7. This is fundamental for a global understanding of the most
  interesting planetary formation mechanisms involved. For example,
  one of the question to be addressed is whether the planetary masses
  depend on the primary stellar masses: the more massive the star, the
  more massive the planets hosted ?
 
  \begin{acknowledgements}

  We acknowledge support from the French CNRS. We are grateful to the
  Observatoire de Haute-Provence (OHP) and to the Programme National de
  Plan\'etologie (PNP, INSU) for the time allocation, and to their
  technical staff.  
  
  These results have made use of the SIMBAD
  database, operated at CDS, Strasbourg, France.
  \end{acknowledgements}


\end{document}